\documentclass[12pt]{article}
\pdfoutput=1
\usepackage{amsmath}
\usepackage{amsfonts}
\usepackage{amssymb}
\usepackage{graphicx}

\usepackage{graphicx,epsfig,sidecap}
\usepackage{bm}
\usepackage{amsfonts,amssymb}
\usepackage{array}
\usepackage{color}
\usepackage{times}
\usepackage{wrapfig}

\def\Tr{{\rm Tr}}
\newcommand{\bx}{{\mathbf x}}
\newcommand{\bC}{{\mathbf C}}
\newcommand{\bU}{{\mathbf U}}
\newcommand{\bG}{{\mathbf G}}
\newcommand{\bT}{{\mathbf T}}
\newcommand{\bV}{{\mathbf V}}
\newcommand{\bI}{{\mathbf I}}
\newcommand{\bN}{{\mathbf N}}
\newcommand{\sech}{{\rm sech}}

\begin{document}
\setcounter{page}{0} \topmargin 0pt
\renewcommand{\thefootnote}{\arabic{footnote}}
\newpage
\setcounter{page}{0}

\begin{titlepage}

\begin{center}
{\Large {\bf R\'{e}nyi information from entropic effects in one higher dimension}}\\

\vspace{2cm}
{\large Mohammad F. Maghrebi\\}  \vspace{0.5cm} {Joint Quantum Institute, and Joint Center for Quantum Information and Computer Science, NIST/University of Maryland, College Park, Maryland 20742, USA}

\vspace{2cm}

\end{center}

\vspace{1cm}

\begin{abstract}
\noindent
  Computing entanglement entropy and its cousins   is often challenging even in the simplest continuum and lattice models, partly because such entropies depend nontrivially on all geometric characteristics of the entangling region.
  Quantum information measures between two or more regions are even more complicated, but contain more, and universal, information. In this paper, we focus on R\'{e}nyi entropy and information of the order $n=2$. For a free field theory, we show that these quantities are mapped to the change of the thermodynamic free energy by introducing boundaries subject to Dirichlet and Neumann boundary conditions in one higher dimension. This mapping allows us to exploit the powerful tools available in the context of thermal Casimir effect, specifically a multipole expansion suited for computing the R\'{e}nyi information between arbitrarily-shaped regions. In particular, we compute the R\'{e}nyi information between two disk-shaped regions at an arbitrary separation distance.
  We provide an alternative representation of the R\'{e}nyi information as a sum over closed-loop polymers, which establishes a connection to purely entropic effects, and proves useful in deriving information inequalities.  Finally, we discuss extensions of our results beyond free field theories.
\end{abstract}

\end{titlepage}

\section{Introduction}
Entanglement is a fundamental property of quantum mechanical states with no counterpart in classical systems.
A widely used measure of entanglement is the von Neumann or entanglement entropy defined as $S(A)=-\Tr(\rho_A \log\rho_A)$ from the reduced density matrix $\rho_A$ of a subsystem $A$, that measures the entanglement of the degrees of freedom in $A$ with those in the rest of the system. This concept has found numerous applications in quantum field theory, black hole thermodynamics, and condensed matter physics. In particular, entanglement entropy exhibits signatures of quantum phase transitions in condensed matter systems \cite{Fazio02,Vidal03,Cardy04,Vedral08}.

Despite its wide range of applications, entanglement (or other types of) entropy is generally hard to compute even in the simplest models. There is still an area of active research aiming to compute various entropies for free lattice models and field theories \cite{Eisler09,Hertzberg11,Safdi12,Cardy13}.
Finding entanglement entropy in certain \emph{critical} models is facilitated by exploiting the techniques of conformal field theory (CFT) \cite{Holzhey94,Cardy09}, but these applications are often limited to 1+1 dimensions.
Part of the challenge lies in the fact that entanglement entropy is defined with respect to a part of a larger system, or a domain in space, and thus depends on its various geometrical properties. This is further complicated if one needs to compute mutual (tripartite, etc.) information between two (three, etc.) regions,
\(
  I(A,B)=S(A)+S(B)-S(A\cup B).
\)
For example, the entanglement entropy of a single interval in a 1+1 dimensional CFT finds a simple logarithmic dependence on the interval's size with a coefficient that only depends on the central charge of the CFT \cite{Holzhey94}. But the mutual information between two disjoint intervals is shown to encode all the information content of the CFT \cite{Furukawa09,Tonni09}.
Powerful geometric representations for entanglement entropy exist in holographic theories, at the expense of a mapping to gravitational models, but nevertheless are limited to CFTs or their massive perturbations \cite{Ryu06,Ryu06-2}.
It is worthwhile to find alternative, simpler geometrical representations that go beyond such limitations. Furthermore, one generally expects that quantum field theories in $d$ dimensions can be mapped to classical ones in $d+1$ dimensions. It is then natural to ask what is the equivalent of entanglement or other types of entropy in one higher dimension. We shall briefly review the standard methods that make use of functional integrals in the $d+1$-dimensional Euclidean space to compute such entropies. But we show that at least in certain cases the mapping to one higher dimension finds a connection with \emph{classical entropic effects}.

In this work, we consider R\'{e}nyi entropy defined as
\(
  R_n(A)=(1-n)^{-1}\log\Tr\left(\rho_A^n\right),
\)
which, for integer $n$, is a simpler alternative to von Neumann entropy, and produces the latter in the limit $n\to 1$.
We exclusively focus on the R\'{e}nyi entropy of the order $n=2$, explicitly defined as
\begin{equation*}
  R_2(A)=-\log\Tr\left(\rho_A^2\right),
\end{equation*}
and study the vacuum/thermal state of a free scalar field theory, but also discuss to what extent our results extend beyond free field theories. It is shown that the R\'{e}nyi entropy $R_2(A)$ of a spatial domain $A$ can be mapped to the change of the thermodynamic free energy by introducing the $d$-dimensional `surface' $A$ in $d+1$ dimensions, where $A$ imposes conventional boundary conditions for a scalar field theory, namely Dirichlet and Neumann boundary conditions. This mapping straightforwardly extends to mutual (tripartite, etc.) R\'{e}nyi information defined as
\begin{equation*}
  I_2(A,B)=R_2(A)+R_2(B)-R_2(A\cup B),
\end{equation*}
and suggests a remarkable connection to \emph{thermal Casimir effect} where an effective interaction energy between two boundaries is induced by thermal fluctuations. Employing the powerful tools---drawn from electrostatics---in the latter context, we show how to compute the R\'{e}nyi information $I_2$ between arbitrarily-shaped regions. Specifically, we compute $I_2$ between two disjoint disk-shaped regions at arbitrary separations, between two parallel half-spaces, and between a half-space and an arbitrary, but small, region. We also exploit an equivalent formulation of the free energy in terms of closed loops (phantom polymers) to show that the  R\'{e}nyi information $I_2$ in $d$ dimensions can be understood from entropic effects (associated with polymer configurations) in $d+1$ dimensions.
This will allow us to easily examine inequalities such as strong subadditivity property.
Furthermore, we extend our results to finite temperature. Finally, we discuss extensions to field theories beyond free theories.

A similar methodology has been adapted to a different, though related, model in Ref.~\cite{Maghrebi15}.
In light of the overlap with Ref.~\cite{Maghrebi15}, we
first give a brief review of the latter, and compare it with our main results here. In the above reference, a model of a scalar field theory coupled to `matter' degrees of freedom (mimicking the electromagnetic field in vacuum coupled to a dielectric material) in $d$ dimensions was considered, and the entanglement between field and matter was studied. It was shown that the mutual information between two material bodies---upon integrating out the scalar field---can be computed from the classical free energy in $d+1$ dimensions where only the scalar field lives in the higher-dimensional space, and is subject to certain (Robin) boundary conditions on $d$-dimensional surfaces of the same shape as the material bodies. Expressing the fluctuating field as a sum over worldlines, or phantom polymers, it was shown that the mutual information finds a geometrical interpretation as a weighted sum over those polymers which intersect the two planar regions in $d+1$ dimensions.
In the present work, we do not consider any matter degrees of freedom, but only a scalar field in vacuum, and study
the quantum information shared between two {\it domains} in space. For the R\'{e}nyi information of the order $n=2$, we find a similar mapping to the free energy in one higher dimension where the $d$-dimensional domains impose different (Dirichlet and Neumann) boundary conditions. Again we find a representation in terms of polymers which are weighted differently than those for the coupled field-matter model of Ref.~\cite{Maghrebi15}. We remark that the mapping to one higher dimension is expected from the quantum-to-classical mapping \cite{Holzhey94,Cardy04}, however, the formulation of the R\'{e}nyi entropy and information as a boundary-value problem subject to the known boundary conditions for the scalar field and the representation of information in terms of a weighted sum over polymers are novel aspects of this work, and allow us to derive a number of quantitative and qualitative results beyond those in the existing literature.

The structure of this paper is as follows. In Sec. \ref{Sec. How to compute}, we briefly review how to compute R\'{e}nyi entropy using functional integral methods. In Sec. \ref{Sec. R2 in quad field theory}, we consider a free field theory at zero/finite temperature, and elaborate in detail on the mapping to thermodynamic and entropic effects in one higher dimension; in Sec. \ref{Sec. I2 from capacitances}, we derive a general expression for the R\'{e}nyi information between two arbitrary regions from the knowledge of their capacitance matrices, which are utilized to compute the R\'{e}nyi information between disks, half-spaces, and small arbitrary regions. We establish a connection to entropic effects in Sec. \ref{Sec. Entropic forces} that enables us to derive information inequalities in Sec. \ref{Sec. Inf inequalities}.
Extension to finite temperature is briefly discussed in Sec. \ref{Sec. Finite T}. We consider specific examples in 1+1-dimensional CFTs beyond free field theories in Sec.~\ref{Sec. Beyond quadratic}. Finally, we summarize our results and discuss future directions in Sec. \ref{Sec. Conclusions}.

\section{How to compute R\'{e}nyi entropy}\label{Sec. How to compute}
\begin{figure}[h]
  \centering
\includegraphics[width=9cm]{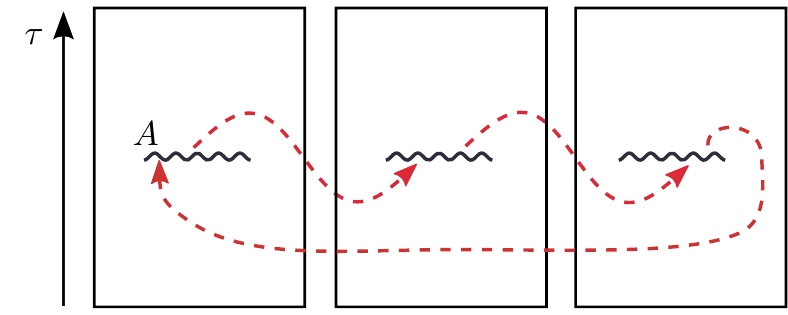}
  \caption{A functional-integral representation for $\Tr\log\left(\rho_A^n\right)$ in an example with $n=3$.} \label{Fig. ThreeCopies}
\end{figure}

Consider a $d+1$-dimensional quantum field theory (in the spatial domain ${\mathbb R}^d$) in its ground state; we shall discuss finite temperature later.
To obtain the R\'{e}nyi entropy, we must compute $\Tr \left(\rho_A^n\right)$; the latter can be computed within a functional-integral formalism which we briefly review in this section \cite{Cardy04,Takayanagi09}.
The wave-functional of the vacuum state of a quantum field theory can be represented by a functional integral from $\tau=-\infty$ to $\tau=0$ as (up to an overall normalization constant)
\begin{equation}
  \Psi\left[\phi_0\right] \propto \int_{\phi(\tau=0,x)=\phi_0(x)} D\phi \, e^{-\int_{\tau=-\infty}^{\tau=0} d\tau L[\phi] },
\end{equation}
with $L$ the Euclidean Lagrangian.
The density matrix of the ground state---a pure state---can be constructed from two copies of the wave functional, $\rho[\phi_0, \phi_0']=\Psi [\phi_0]\bar \Psi[\phi_0']$ where the conjugated functional $\bar\Psi$ can be obtained by a similar functional integral from $\tau=\infty$ to $\tau=0$.
To find the reduced the density matrix $\rho_A$ in a domain $A\subset \mathbb R^d$, we trace out the field outside $A$ by setting $\phi_0(x)=\phi_0'(x)$ over which we integrate,
\begin{equation}\label{Eq. rho from fn integral}
  \rho_A\left[\phi_0, \phi_0'\right] = Z^{-1}\int_{\phi(0^{-},x_A)=\phi_0(x_A),\,\,\phi(0^{+},x_A)=\phi_0'(x_A)}
  D\phi \, e^{-S[\phi] }.
\end{equation}
The action $S=\int_{-\infty}^\infty d\tau L$ is defined over $\tau\in (-\infty,+\infty)$ with a cut across $A$ at time $\tau=0$;
we have used the notation $x_A \in A$.
To normalize the density matrix, the functional integral is divided by $Z$, the vacuum partition function on ${\mathbb R}^{d+1}$. The above representation allows us to cast $\Tr \left(\rho_A^n\right)$ as a functional integral over $n$ copies of space-time:
\begin{equation}\label{Eq. rho to n}
  \hskip -.2in \Tr (\rho_A^n) =Z^{-n} \int_{\phi_{j}(0^-,x_A)=\phi_{j+1}(0^+,x_A)} \,\,\prod_{j=1}^n D\phi_j \, e^{-S[\phi_j]}\equiv Z^{-n} Z_n(A),
\end{equation}
where $n+1 \equiv 1$, and the cut $\{\tau=0\} \times A $ on each copy is glued to neighboring ones as indicated under the integral, see also Fig.~\ref{Fig. ThreeCopies}. $Z_n$ is defined as the functional integral in the above equation.

The R\'{e}nyi entropy is then given by
\begin{equation}\label{Eq. Renyi from part fn}
  R_n(A)=(1-n)^{-1}\log \left[Z^{-n} Z_n(A)\right].
\end{equation}
Similarly, one can obtain the mutual R\'{e}nyi information between the two regions $A$ and $B$.

\section{Free field theories}\label{Sec. R2 in quad field theory}
Consider a scalar field theory $\phi$ that is possibly massive and is subject to an external potential ${\cal V}(x)$, but whose Lagrangian is quadratic,
\begin{equation}\label{Eq. quad Lagrangian}
  L[\phi]=\int d^dx \left[(\partial \phi)^2 + M^2 \phi^2 +{\cal V}(x)\phi^2\right],
\end{equation}
where $\partial^2= \partial_\tau^2+\partial_x^2$; one can view the potential $\cal V$ as a position-dependent correction to the mass squared $M^2$. We shall mostly consider a free massless scalar field theory [$M={\cal V}(x)=0$]. Nevertheless, the main results of this section are equally applicable to a massive field theory in an external potential in an arbitrary dimension. The R\'{e}nyi entropy $R_2$ can be obtained from the functional-integral representation in Eqs.~(\ref{Eq. rho to n}) and (\ref{Eq. Renyi from part fn}),
\begin{equation}\label{Eq. part fn n=2}
  R_2(A)=-\log\left[Z^{-2}\int' D\phi_1 D\phi_2 e^{-S[\phi_1]-S[\phi_2]}\right],
\end{equation}
where the prime on the integral indicates the continuity conditions
\begin{equation*}
  \phi_j(0^-,x)=
    \begin{cases}
    \phi_{j+1}(0^+,x), & x\in A,\\
    \phi_j(0^+,x), & x\notin A,
\end{cases}
\end{equation*}
with $j\in \{1,2\}$ and $j+1\equiv 1$ for $j=2$. Let us define
\begin{equation}
  \phi_\pm=\frac{1}{\sqrt{2}} (\phi_1 \pm \phi_2).
\end{equation}
With this definition, the continuity conditions get decoupled \cite{Casini09,Cardy13}: $\phi_+$ is continuous everywhere across $\tau=0$, while
\begin{equation}\label{Eq. BC for phi-}
  \phi_-(0^-, x)=
  \begin{cases}
    - \phi_-(0^+,x), & x\in A,\\
    \phi_-(0^+,x), & x\notin A.
\end{cases}
\end{equation}
Furthermore, the action being quadratic in the field, we can write
\begin{equation*}
  S[\phi_1]+S[\phi_2]= S[\phi_+]+S[\phi_-].
\end{equation*}
The functional integral in Eq.~(\ref{Eq. part fn n=2}) can be then cast as
\begin{equation}\label{Eq. R2 from phi-}
  R_2(A)=-\log\left[Z^{-1}\int'D\phi_- e^{-S[\phi_-]}\right],
\end{equation}
with the prime indicating the (dis)continuity condition in Eq.~(\ref{Eq. BC for phi-}). The integral over $\phi_+$ being continuous everywhere produces the partition function $Z$, and cancels against one factor of $Z$ outside the integral. The above trick allows us to convert the partition function originally defined on two copies of space-time glued together to a single copy with a cut $A$ at time $\tau=0$ [Eq.~(\ref{Eq. BC for phi-})]. Thus one can view Eq.~(\ref{Eq. R2 from phi-}) as the change of the thermodynamic free-energy in $d+1$ dimensions in the presence of a cut. We stress that this result is more generally valid for all R\'{e}nyi entropies $R_n$ with integer $n$ in free field theories \cite{Casini09}; however, $R_2$ allows even further simplifications as we shall see next.

\begin{figure}[h]
  \centering
 \includegraphics[width=15cm]{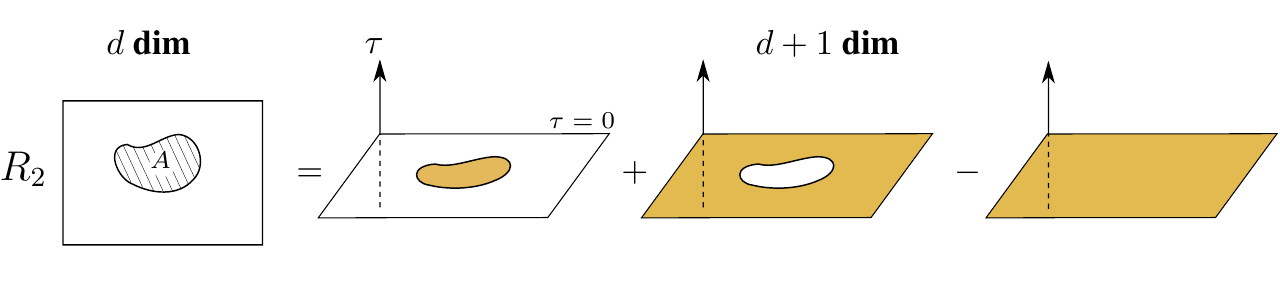}
  \caption{R\'{e}nyi entropy $R_2(A)$ for a region $A$ in $d$ dimensions from the thermodynamic free energy in $d+1$ dimensions. In one higher dimension, $A$ and $\tilde A$ become surfaces of codimension 1 subject to Dirichlet boundary conditions.} \label{Fig. Dirichlet}
\end{figure}

One may hope that the condition (\ref{Eq. BC for phi-}) can be cast in terms of the more familiar boundary conditions allowed for a scalar field theory. Indeed, this is possible by exploiting the (imaginary-)time-reversal symmetry: One can decompose the field $\phi(\tau,x)$ into symmetric and antisymmetric parts in $\tau$, which are defined in the domain $[0,+\infty)\times{\mathbb R}^d $. Extending their domain back to the whole space-time ${\mathbb R}^{d+1}$, one recovers the original partition function with simple boundary conditions on $A$. A simpler alternative is however to break the space of functionals to those even and odd in the imaginary time $\tau$. This is well justified because a quadratic action allows us to write the functional integral as $\exp(- \sum_\alpha \omega_\alpha/2)$, where the sum is over all modes $\alpha$, according to the equipartition theorem. These modes can be cast in a basis in which the corresponding eigenfunctions are either even or odd with respect to $\tau$. Therefore, it suffices to sum over even and odd configurations in the functional integral. The condition (\ref{Eq. BC for phi-}) then becomes (we drop the subscript in $\phi_-$)
\begin{align}
  \begin{cases}
    \phi_e(0,x )=0, & x\in A,\\
    \phi_e(\tau,x ) \mbox{  cont' at  } \tau=0, & x\notin A,
\end{cases}
\end{align}
for even functions, and
\begin{align}\label{Eq. BC for phi o}
  \begin{cases}
    \phi_o(0^+,x )=-\phi_o(0^-,x ), & x\in A,\\
    \phi_o(0,x )=0, & x\notin A,
\end{cases}
\end{align}
for odd functions. A change of variables for odd functions,
\begin{equation}
  \tilde \phi_o(\tau,x)=
  \begin{cases}
    \phi_o(\tau,x ), & \tau>0,\\
    -\phi_o(\tau,x ), & \tau<0,
\end{cases}
\end{equation}
casts Eq.~(\ref{Eq. BC for phi o}) as
\begin{align}
  \begin{cases}
    \tilde\phi_o(\tau,x ) \mbox{  cont' at  } \tau=0, & x\in A,\\
    \tilde \phi_o(0,x )=0, & x\notin A.
\end{cases}
\end{align}
It is now clear that the functions $\phi_e$ and $\tilde \phi_o$ are subject to the familiar Dirichlet boundary conditions on $A$ and its complement $\tilde A=\mathbb R^d\backslash A$, respectively.
The functional integral in Eq.~(\ref{Eq. R2 from phi-}) is then proportional to
\begin{equation*}
  \exp\left[-F{\Big(}\mbox{Dir. on } \{\tau=0\}\times A{\Big)}-F\left(\mbox{Dir. on } \{\tau=0\}\times \tilde A \right)\right],
\end{equation*}
where $F$ is the change of the free-energy of a {\it classical} field theory in the Euclidean $\mathbb R^{d+1}$ due to the presence of a $d$-dimensional surface $A$ subject to the Dirichlet boundary conditions; $k_B T^{(d+1)}=1$ for convenience. Notice that we have dropped the constraint that $\phi_e$ and $\tilde \phi_o$ must be even since these functions are continuous everywhere on the $\tau=0$ plane (in contrast with the function $\phi_o$ in Eq.~(\ref{Eq. BC for phi o})), and thus the odd functions vanish on the whole $\tau=0$ plane, and are insensitive to the region $A$.  To avoid overcounting, we must include an $A$-independent normalization constant. The R\'{e}nyi entropy $R_2$ is then given by [Fig.~\ref{Fig. Dirichlet}]
\begin{align}\label{Eq. R2 from F}
  R_2(A)=
   F{\Big(}\mbox{Dir. on } \{\tau=0\}\times A{\Big)}
  +F{\Big(}\mbox{Dir. on } \{\tau=0\}\times \tilde A{\Big)}
  -F{\Big(}\mbox{Dir. on } \{\tau=0\}\times {\mathbb R}^d{\Big)}.
\end{align}
Clearly, we have $R_2(A=\emptyset)= R_2(A=\mathbb R^d)= 0$.
The above expression is symmetric with respect to $A$ and $\tilde A$ as it should for a system in a pure state which in this case is simply the ground state.

Equation (\ref{Eq. R2 from F}) can also be cast in a way that only refers to the boundary conditions on $A$ and not its complement, which would be more convenient for practical calculations, and especially useful when we discuss R\'{e}nyi information. The trick is to use the Babinet principle that relates solutions subject to the Dirichlet boundary conditions on a plane with a hole to those subject to the Neumann boundary conditions on the hole; in the above context, the perforated plane is given by $\{\tau=0\}\times \tilde A$ while $\{\tau=0\}\times A$ forms the hole, see Fig.~\ref{Fig. Dirichlet}.
One can see this by inspecting even and odd modes separately \cite{Graham05}. In the Dirichlet (Neumann) case, the odd (even) solutions are trivial. But even solutions for the perforated plane with Dirichlet boundary conditions can be constructed from odd solutions for the hole subject to the Neumann boundary conditions. More precisely, we have \cite{Graham05}
\begin{equation*}
  F\left(\mbox{Dir. plane w/ hole }\right)-F\left(\mbox{Complementary Neu. disk}\right)=F\left(\mbox{Entire Dir. plane}\right),
\end{equation*}
where the `disk' denotes the hole-shaped region.
With this identification, the expression (\ref{Eq. R2 from F}) is cast as
\begin{equation}\label{Eq. R2 from Dir + Neu}
  R_2(A)=
   F{\Big(}\mbox{Dir. on } \{\tau=0\}\times A{\Big)}
  +F{\Big(}\mbox{Neu. on } \{\tau=0\}\times A{\Big)}.
\end{equation}
In the present context, the emergence of Dirichlet and Neumann boundary conditions may be attributed to a general argument on the basis of renormalization group (RG) theory.
But we first have to assume that the entropy of the region $A$ can be computed from the partition function in the $d+1$-dimensional Euclidean space with a cut at $A$; this is not obvious as the R\'{e}nyi information of the order $n$ involves $n$ copies of the Euclidean space glued together, but valid for the R\'{e}nyi entropy of the order $n=2$ in the quadratic model above. Nevertheless, given this assumption, one generally needs to compute a quantity of the form
\begin{equation}
  \left\langle e^{-\int_A d^dx \left[\hat V_A\left(\phi(0^+, x)\right)+ \hat V_A\left(\phi(0^-, x)\right)\right]}\right\rangle,
\end{equation}
with a surface potential $\hat V_A$; the average is taken with respect to the Euclidean path integral $\int  D\phi \,e^{-S[\phi]}$ where a cut is assumed across $A$. In the quadratic model described above, the most general function $\hat V_A(\phi)$ respecting the model's inherent $\mathbb Z_2$ symmetry ($\phi \to-\phi$) is simply quadratic,
\begin{equation}
  \hat V_A(\phi)=r\, \phi^2.
\end{equation}
More generally, one can argue that, even in nonlinear field theories, higher orders of $\phi$ are less relevant in the RG sense as the potential $\hat V_A$ lives in one less dimension than the $d+1$-dimensional Euclidean space \cite{Diehl97}. The coefficient $r$ evolves under RG, and goes to one of several fixed points $r=+\infty$, $r=0$, and $r=-\infty$, a familiar result in the context of boundary critical phenomena \cite{Diehl97}. The fixed point $r=+\infty$ makes fluctuations on $A$ infinitely costly, and thus imposes Dirichlet boundary conditions. The fixed point $r=0$ imposes Neumann boundary conditions in the quadratic model considered here, and also in some cases for more general field theories. These boundary conditions are precisely the ones we have encountered in this section. The last fixed point $r=-\infty$ forces the field to diverge on $A$, though it does not arise in the context of our model.

\subsubsection{Area law.}
Let us assume translation invariance, and set ${\cal V}(x)=0$. In the ground state, entanglement and R\'{e}nyi entropies generically exhibit an `area law', that is, the entropy is proportional to the area of the region $A$, although there may be multiplicative logarithmic corrections \cite{Holzhey94,Callan94,Wolf06}; for a review see \cite{Eisert10}.
The area law for the R\'{e}yi entropy $R_2$ can be argued by noting that the free energy $F$ can be expanded as (for a general expansion of the free energy in a magnetic system with boundaries, edges, and corners, see, e.g.,  the introduction in Ref.~\cite{Cardy83})
\begin{equation}
  F{\Big(}\mbox{Dir. on } \{\tau=0\}\times A{\Big)}=  a_0 \,{\rm vol}(A)  +a_1\, {\rm area}(A)+\cdots,
\end{equation}
with $a_i$s possibly cutoff-dependent constants. Note that the vol (area) indicates the volume (area) in $d$ dimensions. With this assumption, one can easily see that the volume contribution drops out from the rhs of Eq.~(\ref{Eq. R2 from F}), and
\begin{equation}
  R_2(A)\sim {\rm area}(A).
\end{equation}
We also point out that similar arguments have appeared elsewhere in the literature \cite{Cardy04,Dubail11,Rajabpour13}.
In $d$ dimensions, the area term goes as $L^{d-1}$ where $L$ is the linear size of the region $A$. In one dimension, the exponent of $L$ goes to zero which may give rise to a logarithmic dependence on $L$ \cite{Callan94,Holzhey94}.

\subsection{Mutual information from capacitance multipoles}\label{Sec. I2 from capacitances}
Here, we will consider the case where $M={\cal V}(x)=0$ for convenience; it is straightforward to generalize to $M> 0$.
While entanglement and R\'{e}nyi entropies are typically cutoff-dependent, the mutual information is free of such ambiguities, and thus provides an attractive, unambiguous alternative to study quantum correlations in quantum field theories.
The mapping to the free energy of a classical field theory in one higher dimension can be similarly carried out for the mutual information. Using Eq.~(\ref{Eq. R2 from Dir + Neu}), the mutual R\'{e}nyi information between the two regions $A$ and $B$ can be cast as
\begin{equation}\label{Eq. I2}
  I_2(A, B)=\sum_{\small\mbox{Dir, Neu}} F(A)+F(B)-F(A\cup B),
\end{equation}
where for the simplicity of notation we have dropped $\{\tau=0\}$ in defining the domains, and made the boundary conditions explicit only in the summation. The above equation also arises in the context of thermal Casimir effect where nearby boundaries attract each other due to thermal (as opposed to quantum) fluctuations. There exist powerful techniques that enable us to compute such effects from electrostatics (or scattering theory for the usual, quantum, Casimir effect).
\begin{figure}[h]
  \centering
  \includegraphics[width=15cm]{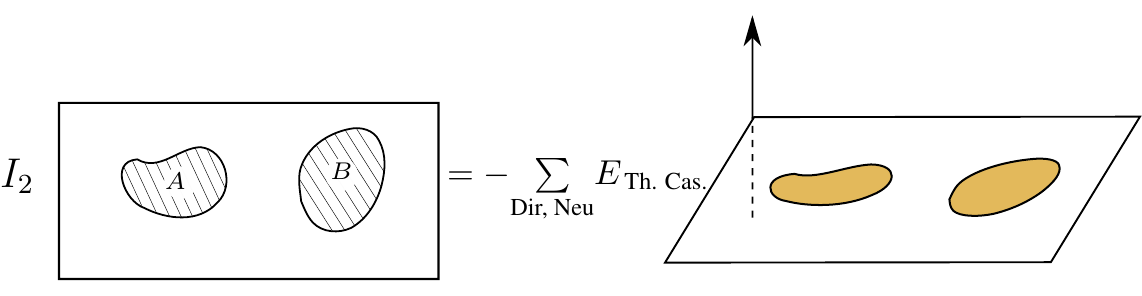}
  \caption{The mutual R\'{e}nyi information $I_2(A,B)$ between the two regions $A$ and $B$ from the thermal Casimir energy $E_{\mbox{\tiny Th.  Cas.}}$ between the disk-like boundaries $A$ and $B$ in one higher dimension. The Casimir energy is summed over Dirichlet and Neumann boundary conditions on the boundaries.} \label{Fig. DirNeu}
\end{figure}
We refer the reader to Refs.~\cite{Kenneth06,Emig08}; here we simply quote the expression that relates the rhs of Eq.~(\ref{Eq. I2}) to the electrostatics of the two disk-like regions $A$ and $B$ immersed in $d+1$ dimensions:
\begin{equation}\label{Eq. CGCG formula}
I_2(A,B)=-\frac{1}{2}\sum_{\small\mbox{Dir, Neu}}\Tr\log\left[\bI- \bC_A \bG \bC_B \bG\right].
\end{equation}
In this equation, $\bI$ is the identity matrix, $\bG$ is the electrostatic Green's function, and $\bC_{A/B}$ denote the capacitance matrices, all defined in $d+1$ dimensions. These matrices can be cast in different bases, the most familiar being the position, $\bx=(\tau, x)$, basis where $\bI(\bx,\bx')={\delta}(\bx-\bx')$,  and $\bG(\bx,\bx')\sim 1/ |\bx-\bx'|^{d-1}$ up to an overall coefficient. The matrix $\bC_A(\bx,\bx')$ depends on the boundary conditions on $A$, the latter generally described by a local `potential' operator $\bV_A(\bx,\bx')=\delta(\bx-\bx') V_A(\bx)$---with the support of $V_A$ on $A$---via the term $\int_\bx V_A(\bx) \phi^2(\bx)$ in the action.
[In the notation defined at the end of Sec.~\ref{Sec. How to compute}, $\hat V_A(\phi(\bx))= V_A(\bx) \, \phi^2(\bx)$.] For example, for Dirichlet boundary conditions, the potential is $V_A(\bx) \to +\infty$ when $\bx \in A$ and zero otherwise.
The matrix $\bC_A$ can be then expressed in terms of the potential operator as $\bC_A= \bV_A \frac{1}{\bI+\bG \bV_A}$ \cite{Kenneth06,Rahi09}. The reader familiar with the Lippmann-Schwinger equation will recognize the similarity to the $\bT$ matrix describing the scattering amplitudes in quantum mechanics. Indeed a similar object describes scattering amplitudes in electrodynamics. We are, however, interested in the equivalent of the $\bT$ matrix for electrostatics---evaluated at zero frequency---hence, the symbol $\bC$ for the capacitance matrix.

It is more convenient to cast the above matrices in a different basis defined by partial waves. For example, for compact objects, angular momentum eigenstates form a natural basis wherein the elements of the $\bC$ matrix give the multipole (monopole, dipole, etc.) response to an external field; these matrix elements depend on the corresponding boundary conditions\footnote{In spherical coordinates in three dimensions $(r,\theta,\phi)$, for example, the capacitance matrix of a compact object is defined as ($Y_l^m$s are spherical harmonics)
\begin{equation*}
  \phi_{lm}=r^l Y_{l}^{m}(\theta,\phi)-\sum_{l'm'}\bC_{l'm',lm}\frac{Y_{l'}^{m'}(\theta,\phi)}{r^{l'+1}},
\end{equation*}
that satisfy the appropriate boundary conditions on the surface of the object.}.
Dirichlet boundary conditions, for example, characterize a perfect conductor. Specifically, the monopole matrix element for the Dirichlet boundary conditions is the usual capacitance.

{\it R\'{e}nyi information at large distances.---}Let us briefly consider the mutual information at a large distance $r$ between two regions in $d>1$ dimensions.
We first note that the logarithm in Eq.~(\ref{Eq. CGCG formula}) can be expanded as $-\Tr\log[\bI-\bN]=\Tr[\bN]+\Tr[\bN^2]/2+\Tr[\bN^3]/3+\cdots$ in a series of multiple reflections \cite{Maghrebi11}.
At large distances between the two regions $A$ and $B$, the leading (single-reflection) term, $(1/2)\Tr\left[\bC_A \bG \bC_B \bG\right]$, suffices.
Moreover, in this regime, Neumann boundary conditions give a negligible contribution compared to the Dirichlet case, and can be dropped (see below and Sec.~\ref{Sec. Entropic forces} for an explanation).
Furthermore, the Green's function $\bG(\bx,\bx')$ varies very little within the domain of each region, thus we have approximately $\bG(\bx,\bx')\sim 1/r^{d-1}$  up to an overall coefficient, independent of $\bx$ and $\bx'$. Therefore, one needs to compute $C^A_0\equiv\langle \phi_0|\bC_A |\phi_0\rangle\equiv\int_\bx\int_{\bx'} \phi^*_0(\bx')\bC_A(\bx',\bx)\phi_0(\bx)$ where $\phi_0(\bx)=\rm const$ characterizes a (suitably normalized) constant function\footnote{The matrix $\bC_A(\bx,\bx')$ has support on $A$.}. In other words, it suffices to keep only the lowest multipole, i.e., the monopole, matrix element $C^A_0$ that characterizes the response to a constant external field. These considerations lead to an expression in terms of the capacitance reported in Ref.~\cite{Cardy13} (see also \cite{Schnitzer14}),
\begin{equation}\label{Eq. Cardy's formula}
  I_2\sim \frac{C^{A}_0 \, C^{B}_0}{2r^{2(d-1)}}.
\end{equation}
An earlier result reporting a similar power-law dependence of the mutual information (corresponding to the von Neumann entropy) at large separations was obtained in Refs.~\cite{Shiba11,Shiba12}.

\begin{figure}[h]
  \centering
  \includegraphics[width=14cm]{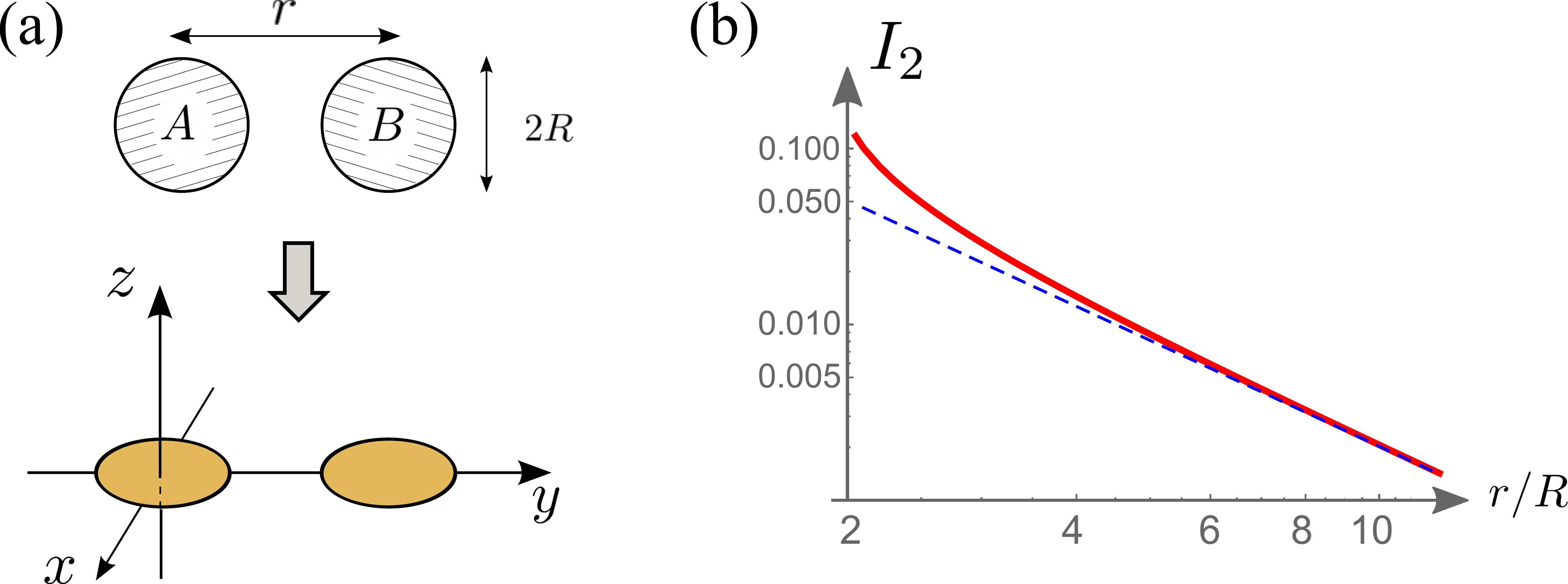}
  \caption{The R\'{e}nyi information $I_2$ between two disk-shaped regions of the radii $R_A=R_B=R$ separated by the distance $r$ in $d=2$ dimensions. (a) The R\'{e}nyi information can be computed from the thermal Casimir energy between two disks embedded on the $z=0$ - plane in $d+1=3$ dimensions. (b) The R\'{e}nyi information $I_2$ as a function of separation is plotted as a thick (red) curve. At large separations, the R\'{e}nyi information falls off as $\sim 2R^2/(\pi^2 r^2)$ plotted as the dashed line.} \label{Fig. TwoDisks}
\end{figure}

{\it R\'{e}nyi information between two disks.---}The R\'{e}nyi information $I_2$ between any two disjoint regions at arbitrary separations can be computed provided that the capacitance matrices of the corresponding regions embedded in one higher dimension are known.  We illustrate this point with the configuration of two circular disks of the radii $R_A=R_B\equiv R$ at the center-to-center separation $r$ in $d=2$ spatial dimensions, see Fig.~\ref{Fig. TwoDisks}. In particular, we obtain the subleading correction to the R\'{e}nyi information at large separations in Eq.~(\ref{Eq. Cardy's formula}).

To find the capacitance matrix of a disk in one higher ($d+1=3$) dimension, it is convenient to work in the oblate spheroidal coordinates $(\xi, \eta, \phi)$ defined as (the coordinate $z$ representing the imaginary time)
\begin{equation}
  x=R\sqrt{(\xi^2+1)(1-\eta^2)}\,\cos\phi, \quad
  y= R\sqrt{(\xi^2+1)(1-\eta^2)}\,\sin\phi, \quad
  z=R \,\xi \eta,
\end{equation}
where $0\le\xi<\infty$, $-1\le\eta\le 1$, and $0\le \phi \le 2\pi$, the latter being the usual azimuthal angle. A disk of the radius $R$ is simply described by the equation $\xi=0$. Also, when $\xi$ is very large, it becomes approximately equal to the radial distance from the center of the disk divided by $R$.  To simplify notation, we take $R=1$ and define all other length scales in units of $R$. Laplacian is separable in spheroidal coordinates in which basis the Green's function [$\bG(\bx,\bx')=1/(4\pi|\bx-\bx'|)]$ can be cast as \cite{Falloon03} (see also \cite{Graham05,Emig09})
\begin{equation}\label{Eq. G < >}
  \bG(\xi,\eta,\phi,\xi',\eta',\phi')=\sum_{n=0}^\infty\sum_{m=-n}^n
  \begin{cases}
    \phi^{\rm reg*}_{nm} (\xi, \eta,\phi) \phi^{\rm out}_{nm}(\xi',\eta',\phi') , & \mbox{if  }\quad \xi <\xi', \\
    \phi^{\rm reg*}_{nm} (\xi',\eta',\phi')  \phi^{\rm out}_{nm} (\xi, \eta,\phi), & \mbox{if  }\quad \xi >\xi',
  \end{cases}
\end{equation}
where the `regular' and `outgoing' functions are defined as
\begin{equation}
  \phi^{\rm reg}_{nm}(\xi,\eta,\phi)={j_n^m}(\xi) Y_{n}^m(\eta,\phi), \qquad  \phi^{\rm out}_{nm}(\xi,\eta,\phi)={h_n^m}(\xi) Y_{n}^m(\eta,\phi).
\end{equation}
In this equation, $Y$s are the usual spherical harmonics, and
the functions $j$ and $h$ are defined as\footnote{We have defined  \begin{align*}
    {j_n^m}(\xi)\equiv i^n \lim_{c\to 0}{R_n^m}^{(1)}(c,-i \xi)/c^n, \quad
    {h_n^m}(\xi)\equiv i^{1-n}\lim_{c\to 0}{R_n^m}^{(2)}(c,-i \xi)\, c^{n+1},
  \end{align*}
  where ${R_n^m}^{(1,2)}$ are the spheroidal functions of the first and the second kinds. The functions $j$ and $h$ are conveniently defined to be real and nonnegative for all $\xi\ge 0$. The Legendre functions in the above definitions should be understood as the third \emph{type} according to {\sc Mathematica} convention for the behavior along the branch cut.}
\begin{align}
  j_n^{m}(\xi)= a_{nm}\, P_n^m(-i\xi ), \qquad h_n^m(\xi)&=b_{nm}\, Q_n^m(-i\xi),
\end{align}
with the coefficients $a_{nm}=\frac{i^n 2^{-2 n-1} (-1)^n \Gamma \left(-n+\frac{1}{2}\right) \Gamma (n-m+1) }{\Gamma \left(n+\frac{3}{2}\right)}$ and $b_{nm}=\frac{i^{1-n} 2^{2 n+1} (-1)^{n+m+1} \Gamma \left(n+\frac{3}{2}\right) }{\Gamma \left(-n+\frac{1}{2}\right) \Gamma (n+m+1)}$.
  Note that the Legnedre function of the first kind $P_n^m(-i \xi)$ is regular everywhere except at infinity, i.e., it diverges as $\xi \to \infty$, while the Legnedre function of the second kind $Q_n^m(-i \xi)$ decays algebraically as $\xi\to\infty$, but has a kink singularity at $\xi=0$. For these reasons, we have designated $\phi^{\rm reg}_{nm}$ and $\phi^{\rm out}_{nm}$ as regular and outgoing functions, respectively (although such terminology is often used for propagating waves).
  In general, the capacitance matrix in the partial-wave basis is related to that of the position basis as $[\bC]_{n'm',nm}=
  \left\langle \phi^{\rm reg}_{n'm'}\!\left|\,\bC \right|\phi^{\rm reg}_{nm}\right\rangle$ where we have used the notation $|\phi\rangle= \int_\bx\phi(\bx)|\bx\rangle$ and $\langle \bx|\bC|\bx'\rangle=\bC(\bx,\bx')$.
  In the above basis, the capacitance matrix is diagonal, $[\bC]_{n'm',nm}=\delta_{n'n}\delta_{m'm} C_{nm}$ where the matrix elements are inferred from the solutions to the Laplace equation
  \begin{equation*}
    \phi_{nm}(\xi,\eta,\phi)= \big[j_n^m(\xi)-C_{nm}\, h_n^m(\xi)\big] Y_n^m(\eta,\phi),
  \end{equation*}
  with the appropriate boundary conditions on the disk at $\xi=0$; we obtain
 \begin{equation}\label{Eq. Capacitance mat}
   C_{nm}^{\rm D}=\frac{{j_n^m}(0)}{{h_n^m}(0)}, \qquad C_{nm}^{\rm N}=\frac{{{j'}_n^m}(0)}{{{h'}_n^m}(0)},
 \end{equation}
  for Dirichlet and Neumann boundary conditions, respectively; the prime indicates the derivative with respect to $\xi$ before setting $\xi=0$.

The Green's functions in Eq.~(\ref{Eq. CGCG formula}) take us between the two disks, and it is advantageous to recast them in a basis that translates the center of coordinates from one disk to another.
Suppose that the Green's function in Eq.~(\ref{Eq. G < >}) is defined with the origin being the center of the disk $A$, and, without loss of generality, take $\xi < \xi'$. We designate $(\xi_{A}, \eta_{A},\phi_{A})$ and $(\xi_{B}, \eta_{B},\phi_{B})$ as the spheroidal coordinates of a given point with respect to the center of the disks $A$ and $B$, respectively (similarly for primed coordinates). We also denote the spheroidal coordinates of the vector that connects the two centers by $(\xi_0, \eta_0,\phi_0)$. Now the outgoing function with respect to $A$ is regular everywhere around the center of $B$, and thus can be expanded in a basis of regular functions with respect to the latter, \begin{equation}
   \phi^{\rm out}_{nm}(\xi'_A,\eta'_A,\phi'_A)=\sum_{n'm'} U_{n'm'nm}(\xi_0, \eta_0,\phi_0)  \phi^{\rm reg}_{n'm'}(\xi'_B,\eta'_B,\phi'_B),
\end{equation}
with the \emph{translation} matrix $U$. The Green's function in Eq.~(\ref{Eq. G < >}) can be then cast as
\begin{equation*}
  \bG(\xi_A, \eta_A,\phi_A, \xi'_B,\eta'_B,\phi'_B)=\sum_{nmn'm'} \phi^{\rm reg}_{n'm'}(\xi'_B,\eta'_B,\phi'_B) U_{n'm'nm}(\xi_0, \eta_0,\phi_0) \phi^{\rm reg *}_{nm}(\xi_A, \eta_A,\phi_A).
\end{equation*}
It is clear from this equation that the Green's function in the partial-wave basis reads $_B\langle \phi^{\rm reg}_{n'm'}|\bG |\phi^{\rm reg}_{nm}\rangle_{A}= U_{n'm'nm}(\xi_0, \eta_0,\phi_0)$ where $|\phi^{\rm reg}\rangle _{A}$ denotes a regular function with respect to $A$ (similarly for $B$).
Following a method similar to Ref.~\cite{Emig08} (see also \cite{Emig09}), we find the translation matrix in the spheroidal basis as
 \begin{align}\label{Eq. Translation mat}
   U&_{n'm'nm}(\xi_0, \eta_0,\phi_0)= (-1)^{n+m}
   \sqrt{4\pi(2n+1)(2n'+1)(2n+2n'+1)}\,\,\times \nonumber \\
   &
   \times\begin{pmatrix}
  n & n' & n+n' \\
  0 & 0 & 0
 \end{pmatrix}
 \begin{pmatrix}
  n & n' & n+n' \\
  m & -m' & -m+m'
 \end{pmatrix}
 h_{n+n'}^{m-m'}(\xi_0) \, Y_{n+n'}^{m-m'}(\eta_0,\phi_0),
 \end{align}
 where the Wigner 3j-symbols are used.

 We are now in a position to compute the expression $\Tr\log(\bI-\bN)$ in Eq.~(\ref{Eq. CGCG formula}). For two identical disks of the radii $R=1$ separated by the distance $r$ along the $y$-axis [Fig.~\ref{Fig. TwoDisks}(a)], we should compute $\bN=\bC {\bU}^+ \bC \bU^-$ separately for Dirichlet and Neumann boundary conditions with the translation matrices
  \(
    [\bU^\pm]_{n'm',nm}=U_{n'm'nm}(\sqrt{r^2-1}\,,0,\pm \pi/2)\equiv U^\pm_{n'm'nm}
  \) being independent of boundary conditions.
  We first compute the leading and subleading terms in the R\'{e}nyi information at large separations. The relevant matrix elements can be deduced from Eqs.~(\ref{Eq. Capacitance mat}) and (\ref{Eq. Translation mat}) as
  \begin{align}
  \begin{split}
    &C^{\rm D}_{00}=\frac{2}{\pi},\qquad C^{\rm D}_{1\, \pm\!1}=\frac{4}{9\pi}, \qquad C^{\rm D}_{10}=0, \\
    & C^{\rm N}_{00}=C^{\rm N}_{1\,\pm\! 1}=0, \qquad \qquad \qquad C^{\rm N}_{10}=-\frac{2}{9\pi},
    \end{split}
  \end{align}
  and
  \begin{align}
  \begin{split}
    &U^\pm_{0000}=\frac{1}{r}+\frac{1}{6 r^3}+{\cal O}\left(\frac{1}{r^5}\right), \quad  U^\pm_{1010}=\frac{3}{r^3}+{\cal O}\left(\frac{1}{r^5}\right),\\
    & U_{001\,\pm\!1}^+=U_{1\,\pm\!1\,00}^+= {U_{001\,\pm\!1}^{-}}^{\!\!\!*}={U_{1\,\pm\!1\,00}^{-}}^{\!\!\!*}=i \sqrt{\frac{3}{2}}\, \frac{1}{r^2} +{\cal O}\left(\frac{1}{r^4}\right),
  \end{split}
  \end{align}
  and, generally, $U^\pm_{n'm'nm}\sim1/r^{n+n'+1}$ at large separations.
Notice that, for Neumann boundary conditions, the monopole matrix element (i.e., the usual capacitance) vanishes, and thus the corresponding contribution to the R\'{e}nyi information decays rather rapidly ($\sim 1/r^6$). We thus focus on the Dirichlet boundary conditions. It is useful to write all the terms that contribute to the leading and subleading terms in the R\'{e}nyi information,
\begin{equation*}\label{Eq. I2 expanded}
  I_2= \frac{1}{2}(C^{\rm D}_{00})^2(U^+_{0000})^2+\frac{1}{4} (C^{\rm D}_{00})^4(U^+_{0000})^4+ 2 C^{\rm D}_{00}C^{\rm D}_{11}|U^+_{0011}|^2+\cdots.
\end{equation*}
The first term on the rhs of the this equation reproduces Eq.~(\ref{Eq. Cardy's formula}) to the leading order ($\sim 2/\pi^2r^2$), but should be also expanded to higher orders to compare against the subleading terms ($\sim 1/r^4$).
The second term comes from the expansion of the $\Tr\log[\bI-\bN]$ to the second power (reflection) of $\bN$, and gives a correction to the subleading term. Finally, the last term on the rhs of the above equation comes from the mixing of monopole and dipole matrix elements. Putting pieces together, we obtain (restoring the units of $R$)
\begin{equation}\label{Eq. I2 leading + sub}
  I_2=\frac{2R^2}{\pi^2 r^2}+\left(\frac{10}{3\pi^2}+\frac{4}{\pi^4}\right)\frac{R^4}{r^4}+{\cal O}\left(({R}/{r})^6\right).
\end{equation}
We also mention that the subleading terms for the mutual information (corresponding to the von Neumann entropy) are computed in Ref.~\cite{Agon15}.

Next we compute the R\'{e}nyi information at all separations between the two disks. To this end, we numerically evaluate the determinant in Eq.~(\ref{Eq. CGCG formula}) by using the exact expression for the capacitance and translation matrix elements in Eqs.~(\ref{Eq. Capacitance mat}) and (\ref{Eq. Translation mat}). The matrix $\bN$ can be truncated to have the dimension $(n+1)^2 \times (n+1)^2$ at the order $n$. At small separations between the two disks, one should include increasingly larger values of $n$. We have considered small separations with the surface-to-surface separation $r/R-2\gtrsim .05$, and confirmed that including `wave numbers' beyond $n=20$ does not change the results by more than a percent (wave numbers up to $n=40$ are examined). The R\'{e}nyi information $I_2$ in a wide range of separations is plotted in Fig.~\ref{Fig. TwoDisks}.
At large separations, $I_2$ falls off as $1/r^2$ given by Eq.~(\ref{Eq. I2 leading + sub}) (or Eq.~(\ref{Eq. Cardy's formula})). To study the short-separation behavior, one needs to numerically evaluate Eq.~(\ref{Eq. CGCG formula}) up to larger values of $n$.

\begin{figure}[h]
  \centering
  \includegraphics[width=12cm]{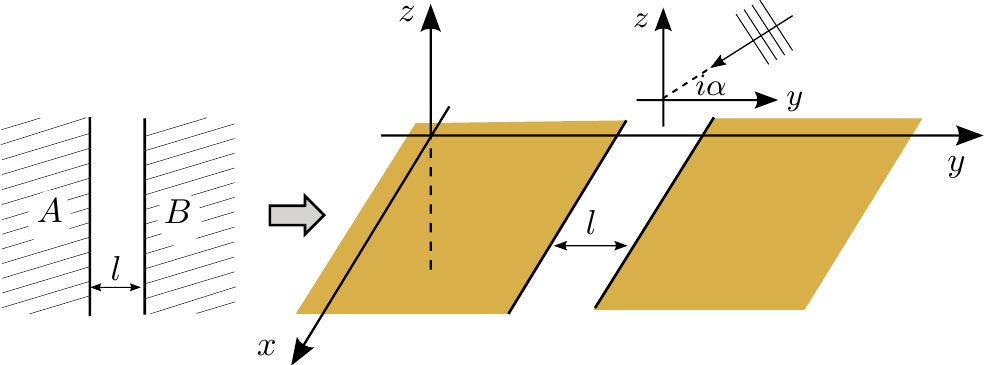}
  \caption{The configuration of two half-spaces separated by $l$ in $d=2$ dimensions, and their mapping to half-plates in $d+1=3$ dimensions. A plane wave is defined by the imaginary angle $i\alpha$ in the $y-z$ plane.} \label{Fig. HalfSpaces}
\end{figure}

{\it R\'{e}nyi information between two half-spaces.---}Next we consider the R\'{e}nyi information $I_2$ between two $d$-dimensional half-spaces whose opposite faces share the same ($d-1$-dimensional) area $\Sigma_{d-1}$, and are separated by the distance $l$.
We first use our intuition of free energy to gain insight into R\'{e}nyi information. In the context of (classical) thermodynamics, a \emph{measurable} quantity is the `force' induced by the separation-dependent free energy.
For the opposite half-spaces, the force should be proportional to the area $\Sigma_{d-1}$, and may depend on the separation $l$. We shall further assume that the force does not depend on the short-wavelength (UV) cutoff, or the regions' extent (IR cutoff) in the normal direction to their surface (as long as it is large compared to $l$). Given that the force is a physical quantity that can be measured, cutoff independence is a reasonable assumption, although a more rigorous argument is desired.
Then, solely on dimensional grounds, the force, $f=-dF(A\cup B)/dl=dI_2(A,B)/dl$, should take a simple scaling form (note that $k_B T^{(d+1)}=1$)
\begin{equation}\label{Eq. f half-spaces}
  f\propto \frac{\Sigma_{d-1}}{l^d}.
\end{equation}
Integrating the force (from infinite separation $l=\infty$ to a finite $l$) to obtain the free energy, and subsequently the R\'{e}nyi information via Eq.~(\ref{Eq. I2}), we find
\begin{align}\label{Eq. area + log}
  I_2&= {\cal A}_d \frac{\Sigma_{d-1}}{l^{d-1}}, \hskip .55in d>1, \\
  \label{Eq. area + log2}&=-{\cal A}_1 \log(l/L), \hskip .22in d=1,
\end{align}
with the coefficients ${\cal A}_d$, and $L$ the extent of the half-\emph{lines} in $d=1$ dimension in which case the boundary is just a point, $\Sigma_0=1$. Note that, for the half-lines of size $L$ in $d=1$ dimension, the power-law in Eq.~(\ref{Eq. f half-spaces}) should be cut off around $l \sim L$ (beyond which the force falls off more rapidly), which leads to Eq.~(\ref{Eq. area + log2}); see Refs.~\cite{Casini04,Furukawa09} for more general results. Therefore, with the assumption that the force does not depend on UV and IR cutoffs, the mutual information will be proportional to the area in $d>1$, while it finds a logarithmic dependence on the regions' size in one dimension \cite{Eisert10}.
The latter is a simple consequence of the scaling form of the force, with the corresponding coefficient ${\cal A}_1=1/4$ known as a special case of a conformal field theory with $c=1$ \cite{Cardy04,Cardy09}.
Below, we compute the corresponding coefficient in $d= 2$ dimensions using electrostatic techniques [Fig.~\ref{Fig. HalfSpaces}]; higher dimensions can be treated similarly.

We first define the cartesian coordinates as the $x$-axis along the edge, the $y$-axis connecting the two half-spaces, and the $z$-axis (in place of $\tau$) perpendicular to the surface, see Fig.~\ref{Fig. HalfSpaces}. Let us consider the capacitance matrix of an individual half-space in a basis defined by the wavevectors $(k_x,k_y,k_z)$. Due to translation symmetry, this matrix is diagonal in $k_x$, hence $\bC_{k_x k_y' k_z', k_x k_y k_z}$; moreover, the laplace equation (for electrostatics) constrains the other two components of the wavevector by $k_x^2+k_y^2+k_z^2=0$. It is useful to represent the wavevector in terms of an imaginary angle $i\alpha$ in the $y-z$ plane,
\begin{equation}
  k_y=i|k_x| \cos(i\alpha), \qquad k_z=i|k_x| \sin (i\alpha).
\end{equation}
In this basis, the capacitance matrix
of the half-space for Dirichlet (D) and Neumann (N) boundary conditions is given by \cite{Maghrebi11}
\begin{align}\label{Eq. Half-space C}
  \bC_{\alpha'k_x, \alpha k_x}^{\rm D/N}=\frac{1}{4\pi}\left[\sech\left(\frac{\alpha+\alpha'}{2}\right)\pm \sech\left(\frac{\alpha-\alpha'}{2}\right)\right],
\end{align}
independent of $k_x$. Next we need to find the free Green's function $\bG$ in Eq.~(\ref{Eq. CGCG formula}) in the above basis. The latter acts as a translation matrix between the two regions separated by the distance $l$. Because of translation symmetry in free space, this matrix is diagonal in \emph{momentum }basis, and is simply given by
\begin{equation}
  \bG_{\alpha'k_x, \alpha k_x}=e^{i k_y l}= e^{-|k_x| \,l\cosh\alpha }.
\end{equation}
The mutual information per unit length of the edge is given by Eq.~(\ref{Eq. CGCG formula}),
\begin{equation*}
     \frac{I_2}{\Sigma_1}=-\frac{1}{2}\sum_{\small\mbox{D, N}}\int_{-\infty}^\infty\frac{dk_x}{2\pi} \int_{-\infty}^\infty \!\!\!d\alpha \, \left[\log\left(\bI- \bC^{\rm D/N} \bG \bC^{\rm D/N} \bG\right)\right]_{\alpha k_x,\alpha k_x}\!\!.
\end{equation*}
To facilitate the computation, we can expand the logarithm in a series of multiple reflections, which, as we shall see, converges rapidly. The leading term in this expansion is (a factor of 2 from restricting the $k_x$ integral to positive values cancels against the overall factor of 1/2)
\begin{align}\label{Eq. MI half-spaces}
  \frac{I_2}{\Sigma_1}{\Large |}_{1^{\rm st} \mbox{\scriptsize ref.}}& =\sum_{\small \mbox{D,N}}\int_0^\infty \frac{dk_x}{2\pi}\int\!\!\int_{-\infty}^\infty d\alpha d\alpha'  e^{-k_x  l \,(\cosh \alpha+\cosh \alpha')} \bC^{\rm D/N}_{\alpha,\alpha'} \bC^{\rm D/N}_{\alpha',\alpha}  \nonumber \\
  &  =\frac{1}{32\pi^3 l} \sum_{\small \mbox{D,N}}\int\!\!\int_{-\infty}^\infty \frac{d\alpha d\alpha'}{\cosh \alpha+\cosh \alpha'} \left[\sech\left(\frac{\alpha+\alpha'}{2}\right)\pm \sech\left(\frac{\alpha-\alpha'}{2}\right)\right]^2 \nonumber \\
  & =\frac{1}{16\pi}\, \frac{1}{l}\,,
\end{align}
which is the sum of the contributions due to both Dirichlet and Neumann boundary conditions. The Neumann's contribution is smaller by an order of magnitude than that of the Dirichlet boundary conditions.
We find that, to the first order in multiple reflections, ${\cal A}_2=1/(16\pi)+\cdots\approx 0.020 +\cdots$ where the ellipses denote corrections due to multiple reflections.
Including the next order in multiple reflections [the second term in the expansion $-\Tr\log(\bI-\bN)=\Tr(\bN)+\Tr(\bN^2)/2+\cdots$], we find
\begin{equation}\label{Eq. A2}
  {\cal A}_2\approx 0.022\,.
\end{equation}
Indeed the corrections at the second order is smaller by an order of magnitude.
In general, the contribution due to the $n$-th order in multiple reflections falls off faster than $1/n^3$ \cite{Maghrebi11}, hence higher orders can be neglected with a good approximation. Furthermore, we remark that similar computations are straightforward in higher dimensions.

The universal coefficients ${\cal A}_d$ can also be calculated by using the entropic $c$-function \cite{Casini05}. In particular, in $d=2$ dimensions, one has
\begin{equation*}
  {\cal A}_2=\frac{1}{\pi} \int_0^\infty \!\! dt \,c_2(t),
\end{equation*}
with the function $c_2$ numerically computed in Ref.~\cite{Casini05}. Using the numerically evaluated function $c_2$ plotted in Fig.~1 of Ref.~\cite{Casini05}, we have estimated the above integral, and found a numerically estimated value of ${\cal A} \approx 0.023$ in reasonable agreement with Eq.~(\ref{Eq. A2}).

\begin{figure}[h]
  \centering
  \includegraphics[width=11cm]{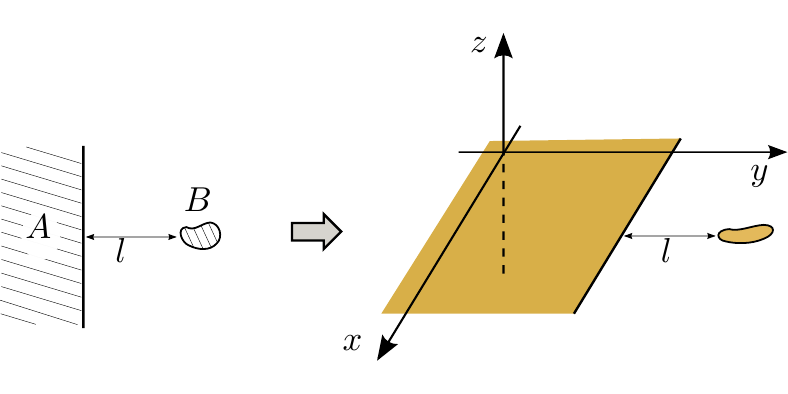}
  \caption{The configuration of a half-space and an arbitrarily-shaped (but small) region in $d=2$ dimensions. The R\'{e}nyi information $I_2$ between the two regions is proportional to the capacitance of a surface of the shape $B$ immersed in $d+1=3$ dimensions.} \label{Fig. Halfplate-disk}
\end{figure}

{\it R\'{e}nyi information between a half-space and a small disk.---}As another example, we consider the R\'{e}nyi information between a half-space and an arbitrary, but small, region in $d=2$ spatial dimensions, see Fig.~\ref{Fig. Halfplate-disk}. The separation $l$ between the two is assumed to be large compared to the linear size $R$ of the small disk-like region. The R\'{e}nyi information can be obtained by combining our treatment of the disk-like regions at large separations with that of the opposite half-spaces. First, we can restrict ourselves to Dirichlet boundary conditions (the contribution due to Neumann boundary conditions appears in higher orders of $R/l$). For the same reason, we shall include only the monopole capacitance element $C_0$ of the small region, and expand the $\Tr\log[1-\bN]$ up to the first power of $\bN$. The calculation proceeds along the same lines as the one described above: Just replace one factor of the capacitance matrix of the half-plate in Eq.~(\ref{Eq. MI half-spaces}) by the capacitance (monopole) $C_0$ of the disk-like region; we find
\begin{align}
I_2
&= \int_0^\infty \frac{dk_x}{2\pi}\int\!\!\int_{-\infty}^\infty d\alpha d\alpha'  e^{-k_x  l \,(\cosh \alpha+\cosh \alpha')} \bC^{\rm D}_{\alpha',\alpha} \,C_0+{\cal O}\left((R/l)^2\right) \nonumber \\
&=\frac{C_0}{8\pi^2 l}\int\!\!\int_{-\infty}^\infty \frac{d\alpha d\alpha'}{\cosh \alpha +\cosh \alpha'} \left[\sech\left(\frac{\alpha+\alpha'}{2}\right)+ \sech\left(\frac{\alpha-\alpha'}{2}\right)\right] +{\cal O}\left((R/l)^2\right) \nonumber \\
&= \frac{C_0}{2\pi l}+{\cal O}\left((R/l)^2\right).
\end{align}
[The capacitance matrix $\bC_{\alpha,\alpha'}$ for the half-plate in three dimensions is given by Eq.~(\ref{Eq. Half-space C}).] Note that, in $d+1=3$ dimensions, the capacitance $C_0$ has the dimension of a length scale, and, for a region of the linear size $R$, we generally have $C_0 \sim R$. For a disk-shaped region of the radius $R$, the capacitance is $C_0=2R/\pi$, and the mutual information between the disk and the half-space is given by
\begin{equation}
  I_2=\frac{1}{\pi^2}\frac{R}{l} +{\cal O}(({R}/{l})^2).
\end{equation}

\begin{figure}[h]
  \centering
  \includegraphics[width=8cm]{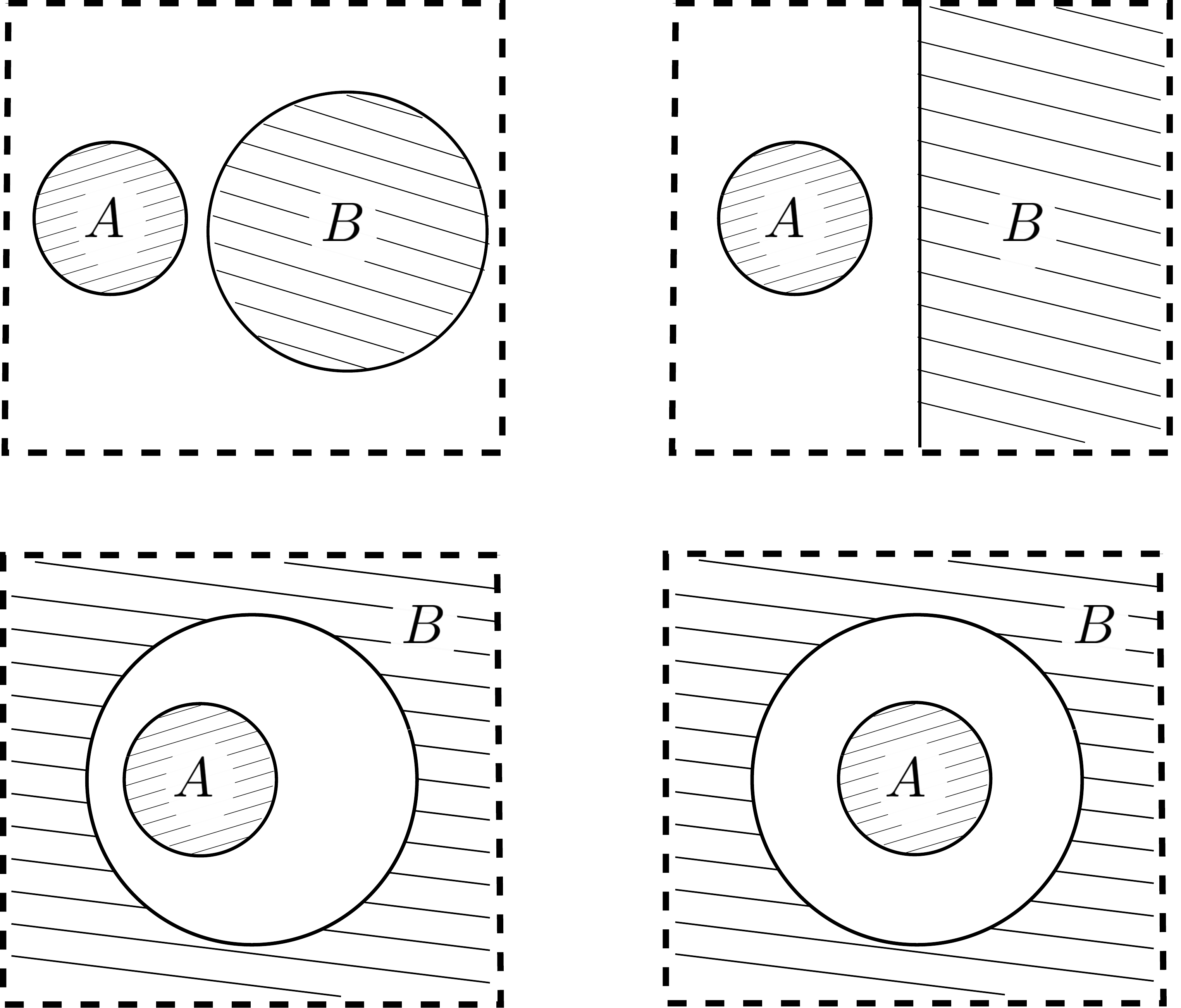}
  \caption{The R\'{e}nyi information $I_2$ can be computed,  using similar techniques, for the above geometries: Two disks with different radii ($R_A\ne R_B$), a disk and a half-space at arbitrary separation ($R_B=\infty$), a disk inside a hole (the radius of curvature $R_B<0$), a disk at the hole center forming an annulus.} \label{Fig. Others}
\end{figure}

{\it R\'{e}nyi information in more general configurations.---}Finally, we remark that the methods and techniques described in this section can be immediately applied to many other interesting cases a few examples of which are illustrated in Fig.~\ref{Fig. Others}. Generalizations to higher dimensions as well as (inner or outer) boundaries of elliptical (rather than circular) shapes are straightforward. In fact, in $d+1=3$ dimensions, Laplacian is separable in 11 coordinate systems \cite{Morse53} suitable for finding analytical expressions as well as facilitating numerical evaluations in specific geometries.
In general, Casimir energy can be computed by a plethora of methods including those inspired by scattering theory \cite{Kenneth06,Emig08}, the Krein formula \cite{Wirzba06}, the worldline formalism \cite{Gies06}, geometric optics \cite{Jaffe04}, as well as various numerical methods \cite{Reid09,Pasquali08,Rodriguez15}, any one of which would be useful for computing the R\'{e}nyi information $I_2$ in a free field theory.

\subsubsection{Scale-free regions.}
Consider two scale-free regions (for example, two cones) $A$ and $B$ separated by the distance $l$. The R\'{e}nyi information $R_2$ between these regions can be computed from the classical free energy as in Eq.~(\ref{Eq. I2}).
Assuming that the force does not depend on the UV and IR cutoffs, the only length scale in the problem is the separation distance $l$, and thus, on dimensional grounds, one finds a simple scaling form for the force as ($k_B T^{(d+1)}=1$)
\begin{equation}
  f=-a_{A,B}\frac{1}{l}\,,
\end{equation}
with the coefficient $a_{A,B}$ depending on both $A$ and $B$ and their relative orientation. Similar arguments have appeared in the context of statistical physics \cite{Binder82,Duplantier86,Maghrebi12}.
The R\'{e}nyi information can be obtained by integrating the force from $M^{-1}$ to $l$; a small mass $M$ is assumed that provides an (IR) cutoff $M^{-1}$ beyond which correlation functions decay exponentially, but alternatively another cutoff such as the system size may be used. Hence, the R\'{e}nyi information finds a logarithmic dependence on separation (up to an additive constant term)
\begin{equation}\label{Eq. log M l}
  I_2=-a_{A,B}\log(M \,l )\,.
\end{equation}
Therefore, a logarithmic term appears for the mutual information between scale-free regions at criticality where $M\to 0$. While this conclusion is strictly based on the R\'{e}nyi information of the order $n=2$ in a free field theory, it is reasonable to expect that it should hold more generally.

Now the total `work' $W\equiv \Delta F(A\cup B) =-\Delta I_2(A,B)$ done in bringing the two `objects' from a tip-to-tip separation $\epsilon$, a short-wavelength cutoff, to infinite separation (or any separation distance $\gg M^{-1}$) is given by
\begin{equation}
  W =-a_{A,B}\log (M \,\epsilon ).
\end{equation}

On the other hand, in even spatial dimensions, a domain with a sharp corner leads to a universal logarithmic term in the (entanglement or R\'{e}nyi) entropy that depends on the (corner-like) singularity ${\cal C}_A$ of the corresponding domain $A$ \cite{Casini07},
\begin{equation}\label{Eq. log}
 R_n(A){\big|}_{\log}=\sigma_n\left({\cal C}_A\right) \, \log (M\, \epsilon).
\end{equation}
\begin{figure}[h]
  \centering
  \includegraphics[width=11cm]{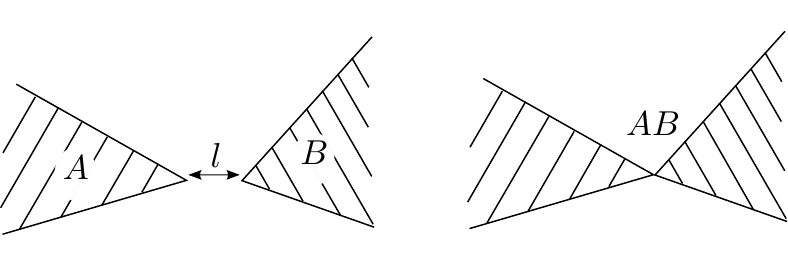}
  \caption{Two-dimensional scale-free regions $A$ and $B$ with sharp corners separated by $l$ or touching at the singular points. The mutual information $I_2(A,B)$ finds a logarithmic form with a universal coefficient that is related to the universal coefficients associated with the sharp corners of $A$, $B$, and $AB$.} \label{Fig. Corners}
\end{figure}
Notice that when the two regions merge, a new region $AB$ forms with a point-like singularity which we denote by ${\cal C}_{AB}$. The work $W$ done in bringing the two objects together is then given by
\begin{align}
  W=-\left[\sigma_2\left({\cal C}_{AB}\right)-\sigma_2\left({\cal C}_A\right) -\sigma_2\left({\cal C}_B\right)\right]\log(M\, \epsilon)\,.
\end{align}
In deriving this equation, we have used the fact that all the leading terms preceding the logarithm in Eq.~(\ref{Eq. log}) are local functions on the boundary \cite{Casini07}, and are independent of corners, and thus cancel out in the difference of the free energy from a configuration where $A$ and $B$ are far separated to one where the two are joined. A comparison of the above expressions for the work yields
\begin{equation}\label{Eq. a from C}
  a_{A,B}=\sigma_2\left({\cal C}_{AB}\right)-\sigma_2\left({\cal C}_A\right) -\sigma_2\left({\cal C}_B\right).
\end{equation}
Interestingly, this equation determines the coefficient of the mutual information between two scale-free regions in even spatial dimensions in terms of the universal coefficients describing the corner contribution to the entropy for individual regions $A$, $B$, and $AB$.
While the universal coefficient for a simple corner in 2+1 dimensions is known \cite{Casini07}, we are not aware of similar expressions for more complicated corner-like structures similar to $AB$ in Fig.~\ref{Fig. Corners}.

\subsubsection{Monotonicity of mutual information.}
It is natural to expect that the mutual information between two regions is larger at shorter separation. In fact, one would generically (excluding contrived situations) expect that the mutual information between two regions of a given shape and orientation increases monotonically with decreasing their separation distance. Drawing the analogy with the thermal Casimir energy for the R\'{e}nyi information $I_2$, this means that the force between two bodies (located on the plane $\tau=0$), subject to Dirichlet and Neumann boundary conditions, should be always attractive! Indeed general theorems have been proved showing that two bodies related by mirror symmetry always attract each other irrespective of their shape and material properties \cite{Kenneth06,Bachas07}.
Specifically, such proofs include Dirichlet boundary conditions in a special limit, while the case of Neumann boundary conditions has so far evaded a rigorous proof although it is also expected to lead to attractive forces.
\begin{figure}[h]
  \centering
\includegraphics[width=6cm]{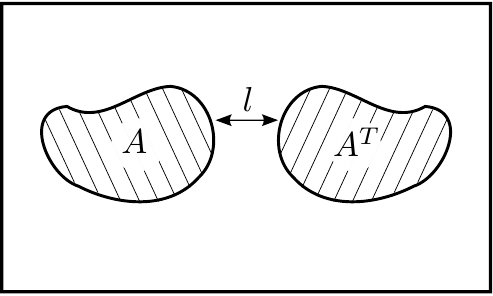}
  \caption{Mutual R\'{e}nyi information between mirror-imaged regions.}
  \label{Fig. Opposites}
\end{figure}
Assuming that a similar conclusion holds for Neumann boundary conditions, it follows that the mutual information $I_2$ increases monotonically with decreasing the separation $l$,
\begin{equation}\label{Eq. monotonic}
  f=\frac{d}{dl}I_2(A, A^T)<0,
\end{equation}
where $A^T$ denotes the mirror image of the region $A$ [Fig.~\ref{Fig. Opposites}].
It is worthwhile to show similar monotonic behaviors for R\'{e}nyi information with $n \ne 2$ in such models.

\subsection{Entropic forces}\label{Sec. Entropic forces}
\begin{figure}[h]
  \centering
  \includegraphics[width=7cm]{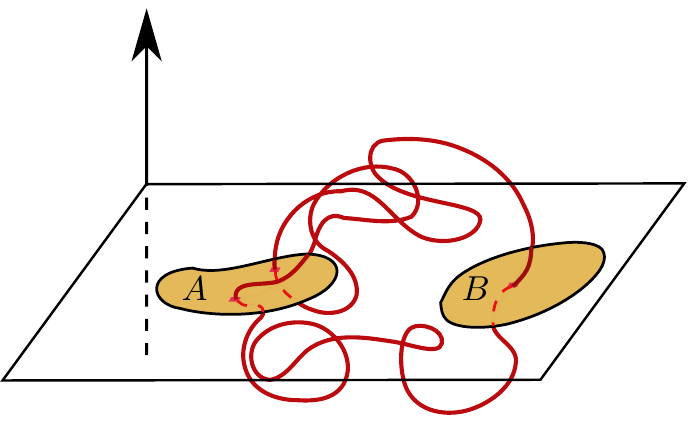}
  \caption{The mutual information $I_2(A,B)$ from a sum over closed-loop polymers. Polymers that intersect both $A$ and $B$ contribute to the mutual information through the Dirichlet boundary conditions, while, for Neumann boundary conditions, they should intersect $A$ and $B$, but not $\mathbb R^d\backslash (A\cup B)$.}
  \label{Fig. Polymer}
\end{figure}

In this section, we shall assume $M={\cal V}(x)=0$ although similar results may be obtained when $M>0$. It is well-known that Casimir energy for certain boundary conditions finds an equivalent formulation in terms of phantom (non-self-avoiding) polymers \cite{Gies06}, inspired by the worldline formalism \cite{Schubert01}. Dirichlet and Neumann boundary conditions should be treated separately. The rhs of Eq.~(\ref{Eq. I2}) for Dirichlet boundary conditions is obtained by summing over all closed-loop polymer configurations in $\mathbb R^{d+1}$ that intersect both $d$-dimensional surfaces $A$ and $B$ (we have dropped $\{\tau=0\}$ for simplicity), see Fig.~\ref{Fig. Polymer}.
More precisely, one finds \cite{Gies06}
\begin{align}\label{Eq. polymer two objects}
    \mbox{ Dir.}: F(A)+F(B)-F(A\cup B)\propto \!\!\int \!\!\frac{ds}{s^{\frac{d+3}{2}}}\int \!\!d^{d+1}\tilde x_{\rm CM} \, \big\langle {\cal N}(A)+{\cal N}(B)-{\cal N}(A\cup B) \big\rangle,
\end{align}
where $\tilde x_{\rm CM}$ and $s$ denote the ($d+1$) coordinates of the `center-of-mass' and the length of the polymer, respectively; the bracket $\langle\cdot\rangle$ averages over all polymers with the same $(x_{\rm CM}, s)$, and $\cal N$ is defined as
\begin{align}
{\cal N}(A)=
\begin{cases}
    1, & \mbox{polymer intersects } A,\\
    0, & \mbox{polymer does not intersect } A.
\end{cases}
\end{align}
It is thus clear that a polymer configuration contributes to Eq.~(\ref{Eq. polymer two objects}) if it intersects both surfaces: Introducing a more concise notation, ${{\cal N}}(A,B)\equiv{\cal N}(A)+{\cal N}(B)-{\cal N}(A\cup B) $, we have
\begin{align}\label{Eq. N Dir}
{{\cal N}}(A,B)=
\begin{cases}
    1, & \mbox{polymer intersects both $A$ and $B$},\\
    0, & \mbox{otherwise},
\end{cases}
\end{align}
and the contribution of the Dirichlet boundary conditions to the mutual R\'{e}nyi information is given by
\begin{equation}\label{Eq. polymer two objects Dir}
    I^{\rm Dir}_2(A,B)\propto \int \frac{ds}{s^{\frac{d+3}{2}}}\int d^{d+1} \tilde x_{\rm CM} \, \big\langle {{\cal N}}(A, B) \big\rangle.
\end{equation}
The Neumann boundary conditions can be formulated similarly. In fact, we can transform back to the Dirichlet boundary conditions on the complementary regions $\tilde A$, $\tilde B$, and $\widetilde{A\cup B}$ (see the discussion in the beginning of this section), and repeat the above procedure. We find
\begin{equation}\label{Eq. polymer two objects Neu}
    I^{\rm Neu}_2(A,B)\propto \int \frac{ds}{s^{\frac{d+3}{2}}}\int d^{d+1} \tilde x_{\rm CM} \, \big\langle {{\cal N}'}(A, B) \big\rangle,
\end{equation}
where ${\cal N}'$ is defined as (all the surfaces are meant at $\tau=0$)
\begin{align}\label{Eq. N Neu}
 {{\cal N}'}(A,B)=
\begin{cases}
    1, & \mbox{polymer intersects both $A$ and $B$, but not $\mathbb R^{d} \backslash (A\cup B)$},\\
    0, & \mbox{otherwise}.
\end{cases}
\end{align}
In particular, clearly ${\cal N}'\le {\cal N}$, and thus $I_2^{\rm Neu}\le I_2^{\rm Dir}$. At large distances between the two regions $A$ and $B$, the additional restriction for Neumann boundary conditions is so severe that the corresponding contribution to the R\'{e}nyi information falls off much faster than the Dirichlet case, $I_2^{\rm Neu}\ll I_2^{\rm Dir}$, typically with a different power-law as a function of separation; this is in harmony with our earlier discussion of the R\'{e}nyi information at large distances. The above expressions summing over polymer configurations provide yet another powerful tool to compute the R\'{e}nyi information most suited to numerical evaluation \cite{Gies06}. Furthermore, they provide a new conceptual tool to find properties of the mutual information.
Note that the above expressions for the R\'{e}nyi information simply require a sum over different polymer configurations; there is no energetics associated with polymers. In other words, these expressions are purely entropic, and can be computed merely by enumerating the number of states. In this respect, the R\'{e}nyi information in the ground state of a free scalar field theory is related to the purely \emph{classical} entropy in one higher dimension; a similar result for the entanglement entropy of a single region in 1+1 dimensional CFT is obtained in the classic work of Holzhey {\it et al.} ~\cite{Holzhey94}. The analogy with entropic forces enables us to verify inequalities similar to strong subadditivity property \cite{Lieb73}.

\subsection{Information inequalities}\label{Sec. Inf inequalities}
It is clear from Eq.~(\ref{Eq. CGCG formula}) or Eqs.~(\ref{Eq. polymer two objects Dir}) and (\ref{Eq. polymer two objects Neu}) that the R\'{e}nyi information is positive
\begin{equation}
  I_2(A,B)\ge 0.
\end{equation}
While this is not a general property of the R\'{e}nyi information \cite{Bengtsson2006},
it is satisfied in the ground state of our quadratic model.

To discuss strong subadditivty property, we first define the tripartite mutual information as (in a slight abuse of notation, we use the same symbol $I_2$ though with three arguments specifying the three regions)
\begin{align}\label{Eq: entanglement 3 objs}
    \hskip -.6in I_2(A,B,C)=&R_2(A)+R_2(B)+R_2(C)\cr
    &-R_2(A\cup B)-R_2(A\cup C)-R_2(B\cup C)+R_2(A\cup B\cup C).
\end{align}
One can also find an expression for the tripartite information in terms of capacitance matrices, see Ref.~\cite{Emig08}. However, it finds a particularly simple form as a sum over polymers, almost identical to that of the mutual information with the identification $I_2(A,B)\to I_2(A,B,C)=I_2^{\rm Dir}(A,B,C)+I_2^{\rm Neu}(A,B,C)$ as well as ${\cal N}(A,B)\to {\cal N}(A,B,C)$ and ${\cal N}'(A,B) \to {\cal N}'(A,B,C)$ corresponding to Dirichlet and Neumann boundary conditions, respectively. With these substitutions, Eqs.~(\ref{Eq. polymer two objects Dir}) and (\ref{Eq. polymer two objects Neu}) describe the tripartite information with
\begin{align}\label{Eq. N tri Dir}
{{\cal N}}(A,B,C)=
\begin{cases}
    1, & \mbox{polymer intersects all $A$, $B$ and $C$},\\
    0, & \mbox{otherwise},
\end{cases}
\end{align}
and
\begin{align}\label{Eq. N tri Neu}
{{\cal N}'}(A,B,C)=
\begin{cases}
    -1, & \mbox{polymer intersects all $A$, $B$ and $C$, but not $\mathbb R^{d} \backslash (A\cup B\cup C)$},\\
    0, & \mbox{otherwise}.
\end{cases}
\end{align}
For Dirichlet boundary conditions, the polymer contribution to the tripartite R\'{e}nyi information, ${\cal N}(A,B,C)$, closely resembles that of the mutual R\'{e}nyi information in Eq.~(\ref{Eq. N Dir}), while, for Neumann boundary conditions, ${\cal N}'(A,B,C)$ is different from the obvious generalization of Eq.~(\ref{Eq. N Neu}) by a negative sign; in fact, $I_2^{\rm Neu}(A,B,C)\le0$. Nevertheless, $|I_2^{\rm Neu}(A,B,C)|\le I_2^{\rm Dir}(A,B,C)$ since the condition (\ref{Eq. N tri Neu}) is more restrictive than (\ref{Eq. N tri Dir}), which implies $I_2(A,B,C)\ge0$.
Having expressed the mutual and tripartite R\'{e}nyi information in terms of a sum over polymers, one immediately finds
\begin{equation}\label{Eq. inequality}
  0\le I_2(A,B,C) \le I_2(A,B),
\end{equation}
where the latter inequality simply follows from ${\cal N}(A,B,C)\le {\cal N}(A,B)$, that is, the number of polymers intersecting all three regions $A$, $B$ and $C$ is less than or equal to those intersecting $A$ and $B$; the fact that ${\cal N}'(A,B,C)\le 0$ is in favor of the second inequality in the above equation. The above inequality for the R\'{e}nyi information does not generally hold in an arbitrary quantum system \cite{Bengtsson2006}; the similar-looking inequality for entanglement (as opposed to R\'{e}nyi) entropy forms the well-known strong subadditivity property valid for all quantum systems \cite{Lieb73}. The rather trivial verification of the inequality in Eq.~(\ref{Eq. inequality}) is intimately related to the mapping to the thermodynamics and entropic forces. In fact, more generally, it is shown that the strong subadditivity property is satisfied for the R\'{e}nyi entropy $R_2$ for all Gaussian states \cite{Adesso12}; this follows from the fact that $R_2$ becomes identical to the Shannon mutual information for a Gaussian state, and in particular the ground state of a free field theory \cite{Rajabpour13}. However, we remark that strong subadditivity property for the von Neumann entropy can be verified in the closely related model of Ref.~\cite{Maghrebi15} with a similar methodology.

Finally, we also note that the above inequalities can be generalized to more than three regions.

\subsection{Finite temperature}\label{Sec. Finite T}
In this section, we derive a formal expression for the R\'{e}nyi entropy $R_2$ at a finite temperature $T=1/\beta$.
There are extensive results on entanglement at finite temperature especially in one spatial dimension \cite{Cardy04}.
In this case, the classical field theory in $d+1$ dimensions is defined on a cylinder $S^1 \times \mathbb R^d$ with the time direction along $S^1$ of the length $\beta$. We need to compute the partition function (\ref{Eq. part fn n=2}) with the continuity conditions
\begin{equation}
  \phi_j(\beta,x)=
  \begin{cases}
    \phi_{j+1}(0,x), & x\in A,\\
    \phi_j(0,x), & x\notin A.
\end{cases}
\end{equation}
\begin{figure}[h]
  \centering
  \includegraphics[width=13cm]{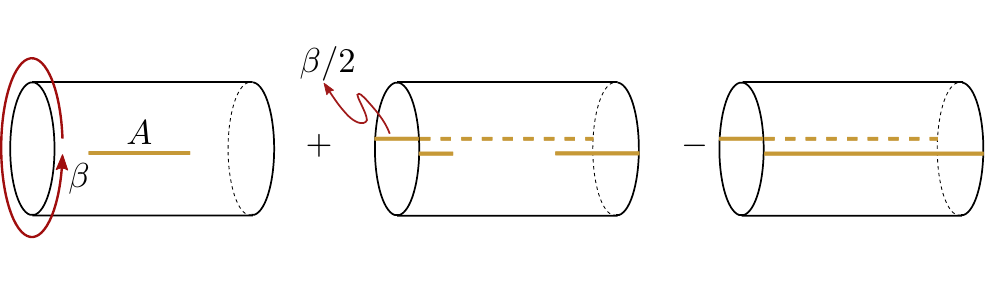}
  \caption{R\'{e}nyi information $R_2(A)$ at a finite temperature $\beta^{-1}$ from the thermodynamic free energy on a cylinder of the circumference $\beta$. The thick lines (surfaces of codimension 1 in $d>1$ dimensions) impose Dirichlet boundary conditions.} \label{Fig. Cylinder}
\end{figure}
Defining the linear combinations $\phi_\pm=(\phi_1\pm\phi_2)/\sqrt{2}$, we again find that $\phi_+$ is continuous everywhere, and whose contribution cancels against one factor of $Z$, while
\begin{equation}
  \phi_-(\beta, x)=
  \begin{cases}
    - \phi_-(0,x), & x\in A,\\
    \phi_-(0,x), & x\notin A.
\end{cases}
\end{equation}
Exploiting the symmetry, we can decompose the space of all functions to those even and odd in time (the mirror point of $\tau$ is given by $\beta-\tau$). The continuity conditions then take the form
\begin{align}
  \begin{cases}
    \phi_e(0,x )=\phi_e(\beta,x )=0, & x\in A,\\
    \phi_e(0,x )=\phi_e(\beta,x ) , & x\notin A,
\end{cases}
\end{align}
for even functions, that is, even functions are continuous along $S^1$, and subject to Dirichlet boundary conditions on $\{\tau=0\}\times A$. For odd functions, we find
\begin{align}
  \begin{cases}
    \phi_o(0,x )=-\phi_o(\beta,x ), & x\in A,\\
    \phi_o(0,x )=\phi_o(\beta,x )=0, & x\notin A.
\end{cases}
\end{align}
To remedy the discontinuity, we make a change of variables
\begin{equation}
  \tilde \phi_o(\tau,x)=
  \begin{cases}
    \phi_o(\tau,x ), & 0<\tau<\beta/2,\\
    -\phi_o(\tau,x ), & \beta/2< \tau<\beta,
\end{cases}
\end{equation}
however, we must note that the original odd functions vanish at $\tau=\beta/2$. Thus the continuity equations in terms of the newly defined variable casts Eq.~(\ref{Eq. BC for phi o}) as
\begin{align}
  \begin{cases}
    \tilde\phi_o(0,x )=\tilde\phi_o(\beta,x ), & x\in A,\\
    \tilde \phi_o(0,x )=\tilde \phi_o(\beta,x )=0, & x\notin A, \\
    \tilde \phi_o(\beta/2,x)= 0, & \forall \, x.
\end{cases}
\end{align}
Thus, the function $\tilde\phi_o$, which is even by construction, is continuous along $S^1$, and subject to Dirichlet boundary conditions on $\{\tau=0\}\times\tilde A$ and $\{\tau=\beta/2\}\times\mathbb R^d$. We can relax the evenness of both functions $\phi_e$ and $\tilde\phi_o$, if we subtract an $A$-independent term. The R\'{e}nyi entropy $R_2$ is then given by [Fig.~\ref{Fig. Cylinder}]
\begin{align}\label{Eq. R2 finite T}
  R_2(A)&  =
   F_\beta{\Big(}\mbox{Dir. on } \{\tau=0\}\times A{\Big)}
  +F_\beta{\Big(}\mbox{Dir. on } \{\tau=0\}\times \tilde A\,\, \cup \,\, \{\tau=\beta/2\}\times \mathbb R^d  {\Big)} \cr
  &\null\hskip .1in
  -F_\beta{\Big(}\mbox{Dir. on } \{\tau=0,\beta/2\}\times \mathbb R^d  {\Big)},
\end{align}
with the subscript $\beta$ making explicit the domain of the field theory along the imaginary time.
In this case, $R_2(A=\emptyset)=0$, while $R_2(A=\mathbb R^d) \propto {\rm vol}( \mathbb R^d)$, the latter because the whole system is entangled with the thermal bath.
The above expression is obviously not symmetric with respect to $A$ and $\tilde A$ as the system is in a thermal state. In the limit of zero temperature, i.e., for $\beta\to\infty$, the boundary condition at $\tau=\beta/2$ can be neglected as it is far from the surface at $\tau=0$, which effectively restores the symmetry. Having cast the entropy in the form of Eq.~(\ref{Eq. R2 finite T}), one can use a number of techniques such as the method of images to systematically compute the R\'{e}nyi entropy at finite temperature, find corrections at small temperature, and derive its high-temperature limit.

\section{Beyond free field theories}\label{Sec. Beyond quadratic}
In the presence of nonlinear interactions (such as $\phi^4$ theory), the simple transformations that allowed us to decouple the two copies of space-time would not work, that is, $S[\phi_1]+S[\phi_2]\ne S[\phi_+]+S[\phi_-]$ for $\phi_\pm=(\phi_1\pm\phi_2)/\sqrt{2}$. Therefore, we may not generally expect relations such as Eq.~(\ref{Eq. I2}) for interacting field theories.
Nevertheless, the free-energy expressions share a general feature with the von Neumann and R\'{e}nyi entropies and mutual information in that they characterize the change of the Euclidean partition function with respect to certain defects in space-time; for R\'{e}nyi entropies, for example, the defects are created by gluing together cuts in several copies of space-time, while the thermodynamic expressions are concerned with the defects in the form of boundary conditions---the particular form of boundary conditions may be argued on a general renormalization-group basis outlined in Sec.~\ref{Sec. R2 in quad field theory}.
Therefore, it is reasonable to expect qualitative agreement between the two \begin{wrapfigure}{r}{0.5\textwidth}
  \centering
\includegraphics[width=4.5cm]{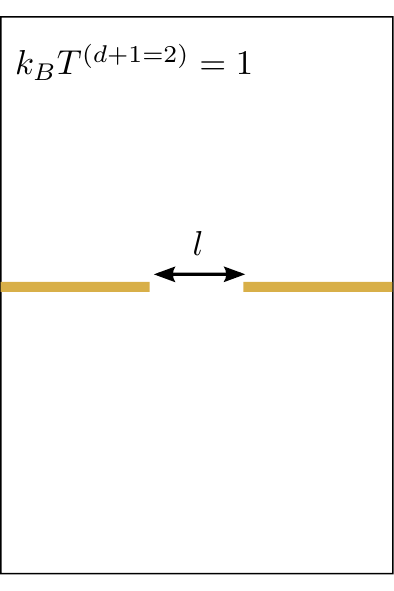}
  \caption{A critical Casimir force between two half-lines is induced by thermal fluctuations in a two-dimensional CFT. } \label{Fig. CriticalCasimir}
\end{wrapfigure}
cases. Below, we demonstrate this point in a specific example of a conformal field theory in 1+1 dimensions. Computing the free energy is fairly complicated for a general interacting field theory, which is also the subject of the critical Casimir effect \cite{Fisher78}; however, CFT techniques in 1+1 dimensions can be utilized to derive exact results for the critical Casimir effect. In particular, we consider two half-lines in one dimension, and compare the two sides of Eq.~(\ref{Eq. I2}), i.e., compare the R\'{e}nyi information of the order $n=2$ against the thermodynamic free energy subject to conformally invariant boundary conditions (Dirichlet and Neumann in the case of the scalar field). We find the agreement between the two expressions, not only in their functional form---which is rather  expected on dimensional grounds---but also in their coefficient. Assuming that the two half-lines are separated by the distance $l$ [Fig.~\ref{Fig. CriticalCasimir}],
we have \cite{Cardy04,Cardy09}
\begin{align}\label{Eq. I2 CFT}
  \frac{d}{dl}I_2(l)= -\frac{c}{4}\frac{1}{l},
\end{align}
while for the force between two half-lines subject to the same conformally-invariant boundary conditions, one finds \cite{Bimonte15} (see also Ref.~\cite{Cardy88})
\begin{align}
  f= -\frac{c}{8}\frac{1}{l}.
\end{align}
The sum of two conformally-invariant boundary conditions (Dirichlet and Neumann boundary conditions for the scalar field, for example) reproduces the same coefficient $c/4$ identical to that of Eq.~(\ref{Eq. I2 CFT}).

As another example, consider two regions  at far separation within a conformal field theory in arbitrary dimensions. A general argument shows that the mutual information decays with the separation $l$ as
\(
   \sim 1/ {l^{2\chi}}
\)
where $\chi$ is the smallest sum of scaling dimensions of operators whose product has the quantum numbers of the vacuum \cite{Cardy13}. The same argument leads to a similar equation for the free energy in the Euclidean space. Whether the corresponding coefficients also agree requires a more elaborate calculation.

Although the precise quantitative agreement may not hold more generally, the above examples show that the Euclidean free energy due to boundary conditions provides a powerful guide to quantum mutual information. For example, one can make similar arguments regarding the area-law in Eq.~(\ref{Eq. area + log}), find a logarithmic law for scale-free regions as in Eq.~(\ref{Eq. log M l}), perhaps even obtain universal relations of the from (\ref{Eq. a from C}), and derive monotonicity relations as in Eq.~(\ref{Eq. monotonic}).

\section{Conclusions and Outlook}\label{Sec. Conclusions}
We have shown that the R\'{e}nyi entropy $R_2(A)$ in the vacuum state of a free quantum field theory maps to the change of the classical free energy in one higher dimension due to the presence of a $d$-dimensional surface subject to Dirichlet and Neumann boundary conditions. Using electrostatic and scattering tools, we have found general expressions relating the mutual R\'{e}nyi information $I_2(A,B)$ in $d$ dimensions to capacitance multipoles of the disk-like regions $A$ and $B$ immersed in $d+1$ dimensions. As a specific example, we have computed the R\'{e}nyi information between two half-spaces, and obtained a result in good agreement with alternative methods based on entropic $c$-functions. While the latter rely on the high symmetry of the regions $A$ and $B$, the multipole expression is completely general, amendable to a number of analytical and numerical treatments. Indeed we have used such techniques to compute the R\'{e}nyi information $I_2$ between two disk-shaped regions at arbitrary separations while only asymptotic expressions---at large separations---are known in the existing literature.

Analogies with the worldline formalism is used to find an alternative representation in terms of closed-loop phantom polymers. The R\'{e}nyi  information then finds a particularly simple interpretation by counting the number of polymers which intersect both $d$-dimensional surfaces $A$ and $B$ in $d+1$ dimensions. In this sense, the R\'{e}nyi information finds a purely entropic form as well as an intriguing geometric picture. Furthermore, it provides a simple way to derive information inequalities. These results can be immediately extended to tripartite (or $m$-partite) information.

Extensions to finite temperature is straightforward, and provides a simple picture for why area laws are expected at zero temperature.
Generalizations to field theories beyond the quadratic model, such as conformal field theories, is far from trivial. We have considered specific examples which however suggest possible connections with our thermodynamic expression.

Future studies would be worthwhile in several directions. First, an extension to von Neumann entropy $S$ and all R\'{e}nyi entropies $R_n$ would be desired. In general, R\'{e}nyi entropies $R_n$ of an integer rank $n>2$ can be constructed similarly by making $n$ copies of space-time, which, for a free field theory, can be decoupled in a suitable basis; however, this procedure would require a complexification of the field when $n>2$, thus complicating the steps described in this work, and possibly giving rise to more complicated boundary conditions. Nevertheless, a similar result for von Neumann entropy exists for the closely related model of Ref.~\cite{Maghrebi15} where one has to sum (integrate) over Robin boundary conditions. Our results are fully applicable to a finite correlation length $M^{-1}$ as well as a finite temperature $T$; further studies in specific examples along these directions would be possible with the general expressions reported in this work. To this end, deriving similar expressions for the mutual information at finite temperature would be desired. Applying the multipole expansion to more than two regions should be interesting, specifically in connection with extensivity \cite{Casini09-2}.
Extensions to fermionic field theories should also be of interest. Finally, it remains to be seen to what extent the conclusions of this work may be extended to interacting quantum field theories or many-body systems.

\section*{Acknowledgements}
I am grateful to Noah Graham for useful discussions regarding spheroidal functions, and to Robert L. Jaffe for his unpublished notes on multiple reflections with capacitance matrices. I also acknowledge useful correspondence with Howard J. Schnitzer. This work was supported by the NSF PIF, AFOSR, ARO, ARL, NSF PFC at the JQI, and AFOSR MURI.

\end{document}